# ON THE SELF-INTERACTION IN THEORY OF REAL DIRAC FIELD

S.Botrić* and K.Ljolje

Academy of Sciences and Arts of Bosnia and Herzegovina,
Sarajevo,Bosnia and Herzegovina

Summary. - Reduction of the self-interaction in the theory of
real Dirac field is considered

*Faculty of Electrical Engineering,Mechanical Engineering and
Naval Architecture,University of Split,Split,Croatia

# 1. INTRODUCTION

In this article we consider reduction of the self-interaction in the real Dirac field theory [1,2]. We rely on the corresponding classical theory [3-7].

# 2. REAL DIRAC FIELD

The real Dirac field interacting with the electromagnetic field is defined by the Lagrangian density [1]

$$L = L_{em} + L_{DI} \qquad (1)$$

where

$$L_{em} = \frac{1}{8\pi}\left(-\frac{1}{2}F_{\alpha\beta}F^{\alpha\beta} + F^2 - G^2\right) \qquad (2)$$

$$L_{DI} = K\left[\frac{\overline{\Psi}(1-a)\Psi}{1-a^2} - \kappa^2 \overline{\Phi}(1+a)\Phi\right] \qquad (3)$$

$$a = \frac{e}{K}A_\beta\eta^\beta, \quad K = mc^2, \quad \kappa = \frac{mc}{\hbar} \qquad (4)$$

$$\Psi = D\Phi, \quad D = \partial_\alpha \eta^\alpha \qquad (5)$$

$$G = \partial_\alpha A^\alpha, \quad F = \partial_\alpha C^\alpha \qquad (6)$$

$A^\alpha$ and $C^\alpha$ are the Lagrange's variables of the electromagnetic field and eight-component matrix $\Phi$ is the Lagrange's variable (variables) of the Dirac field.

The Lagrange's equation for $A^\alpha$ is given by

$$\partial_\mu \partial^\mu A^\alpha = 4\pi e\left[\kappa^2 \overline{\Phi}\eta^\alpha \Phi + \frac{\overline{\Psi}\eta^\alpha \Psi}{1-a^2} - 2\frac{e}{K}\frac{A^\alpha \overline{\Psi}(1-a)\Psi}{(1-a^2)^2}\right] \qquad (7)$$

The canonical equations of the real Dirac field are given by

$$D\Phi - (1+a)\frac{c}{K}\Pi_{\Phi^+} = 0 \qquad (8)$$

$$D\left(\frac{c}{K}\Pi_{\Phi^+}\right) + \kappa^2(1+a)\Phi = 0 \qquad (9)$$

A particular class of solutions one obtains by selection

$$\frac{c}{K}\Pi_{\Phi^+} = \kappa N\Phi \qquad (10)$$

and

$$[D - \kappa(1+a)N]\Phi = 0 \qquad (11)$$

where

$$N = \begin{bmatrix} N_a & 0 \\ 0 & N_b \end{bmatrix}, \quad N_a = \begin{bmatrix} 0 & -1 & 0 & 0 \\ 1 & 0 & 0 & 0 \\ 0 & 0 & 0 & -1 \\ 0 & 0 & 1 & 0 \end{bmatrix}, \quad N_b = \begin{bmatrix} 0 & -1 & 0 & 0 \\ 1 & 0 & 0 & 0 \\ 0 & 0 & 0 & 1 \\ 0 & 0 & -1 & 0 \end{bmatrix}.$$

S-transformation [1] yields

$$\Phi' = S\Phi = \begin{bmatrix} \varphi_a \\ \varphi_b \end{bmatrix}, \tag{12}$$

$$[i\partial_\alpha \gamma^\alpha - \kappa(1+\tilde{a})]\varphi_a = 0, \quad \tilde{a} = \frac{e}{K} A_\beta \gamma^\beta, \tag{13}$$

$$[i\partial_\alpha \gamma^\alpha + \kappa(1+\tilde{a})]\varphi_b = 0, \tag{14}$$

and [1]

$$\varphi_b = N_b \varphi_a^*. \tag{15}$$

For these solutions Eq. (7) becomes

$$\partial_\mu \partial^\mu A^\alpha = 4\pi e \left(2\kappa^2 \overline{\Phi} \eta^\alpha \Phi\right). \tag{16}$$

Equations (16) and (13) determine then the behaviour of the system. Since $A_p^\alpha$ is determined by the Dirac field it contains the self-interaction.

## 3. REDUCTION OF THE SELF-INTERACTION

Effects of the self-interaction we may try to evaluate by the perturbation method and partly include in observable constants similarly to the conventional renormalization procedure. But we prefer here to apply the classical field method. The classical electrodynamics is linear one with respect to the interaction. Due to this reason we consider here also the linear approximation of the Lagrangian (3) with respect to $A^\alpha$, e.g. we take

$$L_{DI} \to L_{DI}^1 = K(\overline{\Psi}\Psi - \kappa^2 \overline{\Phi}\Phi) - \frac{1}{c} A_\alpha j_D^\alpha. \tag{17}$$

where

$$\frac{1}{c} j_D^\alpha = e(\overline{\Psi}\eta^\alpha \Psi + \kappa^2 \overline{\Phi}\eta^\alpha \Phi). \tag{18}$$

and consider only the first order effects.

In the classical theory the self-interaction is reduced to the radiation effects [3-7] by addition of

$$L_s = \frac{1}{2c} A_{D\alpha} j_D^\alpha \quad , \quad A_D^\alpha = \frac{1}{c} \int \frac{1}{|\vec{r} - \vec{r}'|} j_{Dret}^\alpha d^3x' \quad , \tag{19}$$

to the Lagrangian density. Consequently, we do the same taking now only $j_D^\alpha$ given by (18). The new Lagrangian is then*

$$L = L_{em} + L_{DI}^1 + L_s \tag{20}$$

Variations of $S = \int \frac{1}{c} L \, d^4x$ with respect to $A^\alpha$ and $\overline{\Phi}$ yield, respectively,

$$\partial_\mu \partial^\mu A^\alpha = \frac{4\pi}{c} j_D^\alpha \quad , \tag{21}$$

$$D(1 - a + b)\Psi + \kappa^2(1 + a - b)\Phi = 0 \quad , \tag{22}$$

where

$$b = \frac{e}{2K}(A_{Dret\beta} + A_{Dadv\beta})\eta^\beta \equiv \frac{e}{K} B_\beta \eta^\beta \tag{23}$$

In the linear approximation there is also a family of solutions defined by

$$\Psi = \kappa(1 + a - b)N\Phi \quad , \tag{24}$$

$$[D - \kappa(1 + a - b)N]\Phi = 0 \quad . \tag{25}$$

For these solutions

$$\frac{1}{c} j_D^\alpha = e[2\kappa^2 \overline{\Phi}\eta^\alpha \Phi + 2\kappa^2 \frac{e}{K}(A^\alpha - B^\alpha)\overline{\Phi}\Phi + \ldots] \approx 2e\kappa^2 \overline{\Phi}\eta^\alpha \Phi \tag{26}$$

and

$$\partial_\mu \partial^\mu A^\alpha \approx 4\pi e 2\kappa^2 \overline{\Phi}\eta^\alpha \Phi \quad , \quad (\text{with } \partial_\alpha j_D^\alpha \approx 0) \quad . \tag{27}$$

S-transformation yields

$$\Phi' = S\Phi = \begin{bmatrix} \varphi_a \\ \varphi_b \end{bmatrix} , \tag{28}$$

$$[i\partial_\alpha \gamma^\alpha - \kappa(1 + \tilde{a} - \tilde{b})]\varphi_a = 0, \tag{29}$$

$$\varphi_b = N_b \varphi_a * .$$

(30)

Equations (27) and (28-30) determine now behaviour of the system.

General solution of Eq.(27) is given by

---

* It corresponds to the renormalization procedure of the conventional quantum field theory.

$$A^\alpha = A_h^\alpha + A_p^\alpha \quad , \quad A_p^\alpha = A_D^\alpha = \int \frac{1}{|\vec{r} - \vec{r}'|} \left(2\kappa^2 \overline{\Phi} \eta^\alpha \Phi\right)_{ret} d^3x' \ . \tag{31}$$

Substitution of (31) into (25) yields

$$[D - \kappa(1 + a_h + a_{rad})N]\Phi = 0, \tag{32}$$

where

$$a_{rad} = \frac{e}{2K}(A_{Dret\beta} - A_{Dadv\beta})\eta^\beta . \tag{33}$$

Consequently, the self-interaction is reduced to the radiation effects. In absence of the homogeneous solution $A_h^\alpha$ (=0) there exists "free particle stationary states".

Energy of the field

(a) Starting with

$$T_\alpha{}^\beta = \partial_\alpha \chi \frac{\partial L}{\partial(\partial_\beta \chi)} - \delta_\alpha{}^\beta L \ , \tag{34}$$

and applying it to (1) one gets

$$T_\alpha{}^\beta = T_\alpha{}^\beta{}_{em} + T_\alpha{}^\beta{}_{DI} \ , \tag{35}$$

where

$$T_\alpha{}^\beta{}_{em} = \frac{1}{4\pi}\left[-\partial_\alpha A_\gamma F^{\beta\gamma} + \frac{1}{4}\delta_\alpha{}^\beta F_{\mu\nu} F^{\mu\nu}\right], \tag{36}$$

$$T_\alpha{}^\beta{}_{DI} = K\left[\partial_\alpha \overline{\Phi}\eta^\beta \frac{c}{K}\Pi_{\Phi^+} + \frac{c}{K}\Pi_\Phi \eta^o \eta^\beta \partial_\alpha \Phi - \delta_\alpha{}^\beta L_{DI}\right], \tag{37}$$

(here we have taken $C^\alpha = 0$). Using these equations one obtains

$$\partial_\beta T_{em}^{\alpha\beta} = -\frac{1}{c}\partial^\alpha A^\gamma j_{D\gamma}, \tag{38}$$

$$\partial_\beta T_{DI}^\beta = \frac{1}{c}\partial^\alpha A^\gamma j_{D\gamma}, , \tag{39}$$

and

$$\partial_\beta T^{\alpha\beta} = 0. \tag{40}$$

From here follows the energy-momentum constant of motion

$$cP^\alpha = \int \left(T_{em}^{\alpha o} + T_{DI}^{\alpha o}\right)d^3x . \tag{41}$$

The tensor $T_{DI}^{\alpha\beta}$ for the solutions (10-11) takes the form

$$T_{DI}^{\alpha\beta} = K\kappa\left(\partial^\alpha \overline{\Phi}\eta^\beta N\Phi - \overline{\Phi}\eta^\beta N\partial^\alpha \Phi\right). \tag{42}$$

Starting with this tensor by definition and applying it to the equation of motion (11) one gets

$$\partial_\beta T_{DI}^{\alpha\beta} = 2\kappa^2 K\overline{\Phi}(\partial^\alpha a)\Phi =$$

$$= \frac{1}{c}\partial^\alpha A^\gamma j_{D\gamma} . \qquad (43)$$

(b) Equation (43) we apply to the Lagrangian (20). It is necessary only to substitute $a$ with $a-b$ :

$$\partial_\beta T_{DI}^{\alpha\beta} = 2\kappa^2 K\overline{\Phi}(\partial^\alpha a - \partial^\alpha b)\Phi = \frac{1}{c}\partial^\alpha A^\gamma j_{D\gamma} - \frac{1}{c}\partial^\alpha B^\gamma j_{D\gamma} . \qquad (44)$$

Sum of (38) and (44) yields

$$\partial_\beta (T_{em}^{\alpha\beta} + T_{DI}^{\alpha\beta}) = -\frac{1}{c}\partial^\alpha B^\gamma j_{D\gamma} . \qquad (45)$$

Let us denote by "B" the electromagnetic field belonging to $B^\alpha$. Then one gets

$$\partial_\beta T_{em\,B}^{\alpha\beta} = -\frac{1}{c}\partial^\alpha B^\gamma j_{D\gamma} . \qquad (46)$$

Substitution of (46) into (45) yields

$$\partial_\beta (T_{em}^{\alpha\beta} - T_{em\,B}^{\alpha\beta} + T_{DI}^{\alpha\beta}) = 0 . \qquad (47)$$

From here follows the energy-momentum constant of motion

$$cP^\alpha = \int (T_{em}^{\alpha o} - T_{em\,B}^{\alpha o} + T_{DI}^{\alpha o})d^3x . \qquad (48)$$

## 4. HYDROGEN ATOM

Let an external field be given by a point charge (proton) at rest. Explicitely,

$$\frac{1}{c}j_{ext}^\alpha = (|e|\delta(\vec{r}),0) . \qquad (49)$$

Additional term in the Lagrangian density (20) is then

$$L_{ext} = -\frac{1}{c}A_\alpha j_{ext}^\alpha . \qquad (50)$$

This term changes the equation for $A^\alpha$ and $T^{\alpha\beta}$. They become, respectively,

$$\partial_\mu \partial^\beta A^\alpha = \frac{4\pi}{c}(j_D^\alpha + j_{ext}^\alpha) , \qquad (51)$$

$$\partial_\beta (T_{em}^{\alpha\beta} - T_{em\,B}^{\alpha\beta} + T_{DI}^{\alpha\beta} + g^{\alpha\beta}A_\gamma \frac{1}{c}j_{ext}^\gamma) = 0 . \qquad (52)$$

In accordance to (49) the equation (52) becomes

$$\partial_\beta(T_{em}^{\alpha\beta} - T_{em\,B}^{\alpha\beta} + T_{DI}^{\alpha\beta} + g^{\alpha\beta}A_o \frac{1}{c} j_{ext}^o) = 0. \tag{53}$$

The energy-momentum constant of motion is then

$$cP^\alpha = \int (T_{em}^{\alpha o} - T_{em\,B}^{\alpha o} + T_{DI}^{\alpha o} + g^{\alpha o}A_o \frac{1}{c} j_{ext}^o) d^3x. \tag{54}$$

The energy is given by

$$cP^o = \int (T_{em}^{oo} - T_{em\,B}^{oo} + T_{DI}^{oo} + A_o \frac{1}{c} j_{ext}^o) d^3x. \tag{55}$$

The particular solution of Eq.(51) is now given by

$$A_p^\alpha = A_{Dret}^\alpha + A_{ext\,ret}^\alpha. \tag{56}$$

It changes $a-b$ in Eq.(32) into

$$a - b = \frac{e}{K}\left(\frac{|e|}{r}\eta^o + A_{rad\beta}\eta^\beta\right), \tag{57}$$

and for $a_h = 0$ the equation for the real Dirac field becomes

$$\left[D - \kappa\left(1 + \frac{e|e|}{Kr}\eta^o + \frac{e}{K}A_{rad\beta}\eta^\beta\right)N\right]\Phi = 0, \tag{58}$$

and

$$\Phi = S^+\Phi' \equiv S^+\begin{bmatrix}\varphi_a \\ \varphi_b\end{bmatrix}, \quad \varphi_b = N_b\varphi_a*, \tag{59}$$

$$\left[i\partial_\alpha\gamma^\alpha - \kappa\left(1 + \frac{e|e|}{Kr}\gamma^o + \frac{e}{K}A_{rad\beta}\gamma^\beta\right)\right]\varphi_a = 0, \tag{60}$$

There exist stationary solutions of Eq.(60)

$$\varphi_a(x^o,\vec{r}) = e^{-ik_o x^o}\varphi_{ak}(\vec{r}), \tag{61}$$

(for which $A_{rad}^\alpha = 0$). The energy of these states are equal

$$cP^o = \int T_{DI}^{oo}d^3x = K\kappa\int(\partial^o\Phi^+ N\Phi - \Phi^+\partial^o N\Phi)d^3x = \hbar k^o. \tag{62}$$

These energies are exactly equal to those of the conventional relativistic (Dirac) hydrogen atom.

## 5. CONCLUSIONS

The method of reduction of the self-interaction developed in the classical field theory [3-7] is also applicable in the real Dirac field theory.By this the real Dirac field theory becomes closed with full internal consistence.

Application to the hydrogen atom (in the linear electrodynamics) yields observable energies,but they are now obtained quite differently from the conventional quantum theory.It also supports the real Dirac field theory.Thus it tells us more about the nature of physical reality.

References :

[1] K.Ljolje , The Dirac Field in Real Domain,
[2] S.Botrić , Note on the Dirac Field in Real Domain,
[3] L.Lopes and M.Schönberg , Phys. Rev., 67, 122 (1945).
[4] M.Schönberg , Phys. Rev., 69, 211 (1946).
[5] S.N.Gupta, Proc.Phys.Soc. A, 64, 50 (1951)
[6] F.Rohrlich , Classical Charged Particles , Addison-Wesley Publ. Company ,
             Reading , Mass. (1964).
[7] S.Botrić and K.Ljolje , Il Nuovo Cimento , 107B, 1, 51 (1992).